\documentclass[Crystals,review,accept,moreauthors,dvipdfm,12pt,a4paper]{mdpi} 
\lastpage{x}
\doinum{10.3390/------}
\pubvolume{3}
\pubyear{2013}
\history{Received: 16 November 2012; in revised form: 13 December 2012 / Accepted: 9 January 2013 / \linebreak Published: xx}

\usepackage{graphicx}
\usepackage{amssymb}
\usepackage{amsmath}
\usepackage{graphicx}
\usepackage{flafter}
\usepackage{array}
\usepackage{color}
\usepackage{multirow}
\Title{Indirect Exchange and Ruderman-Kittel-Kasuya-Yosida (RKKY) Interactions in Magnetically-Doped Graphene}

\Author{Stephen R. Power$^{1,2,}$* and Mauro S. Ferreira$^{1, 3}$}

\address{%
$^{1}$ School of Physics, Trinity College Dublin, Dublin 2, Ireland; E-Mail: ferreirm@tcd.ie\\
$^{2}$ Center for Nanostructured Graphene (CNG), DTU Nanotech, Department of Micro- and Nanotechnology, Technical University of Denmark, Kongens Lyngby 2800, Denmark\\
$^{3}$ Centre for Research on Adaptive Nanostructures and Nanodevices (CRANN), Trinity College Dublin, Dublin 2, Ireland}

\corres{E-Mail: spow@nanotech.dtu.dk; \linebreak Tel.: +45-4525-5694.}

\abstract{Magnetically-doped graphene systems are potential candidates for application in future spintronic devices. A key step is to understand the pairwise interactions between magnetic impurities embedded in graphene that are mediated by the graphene conduction electrons. A large number of studies have been undertaken to investigate the indirect exchange, or RKKY (Ruderman-Kittel-Kasuya-Yosida), interactions in graphene. Many of these studies report a decay rate faster than expected for a two-dimensional material and the absence of the usual distance dependent oscillations. In this review we summarize the techniques used to calculate the interaction and present the key results obtained to date. The effects of more detailed parameterisations of the magnetic impurities and graphene host are considered, as are results obtained from \emph{ab initio} calculations. Since the fast decay of the interaction presents an obstacle to spintronic applications, we focus in particular on the possibility of augmenting the
interaction range by a number of methods including doping, spin precession and the application of strain.}



\begin{document}

\section{Introduction}
\label{intro}

Graphene has been in the scientific limelight for the past few years due to a range of interesting properties that it has been shown, or predicted, to display. Superlative mechanical properties and a unique set of potentially tunable electronic and optical properties suggest a large number of possible applications for graphene and related materials \cite{riseofgraphene, geim_graphene:_2009, neto:graphrmp,novoselov_nobel_2011}. Apart from potential roles in nanoelectronics, optoelectronics and as reinforcements in composites, graphene-based materials have also been mooted for application in spintronics \cite{yazyev:review}. This field is projected to play a very important role in future technologies as it may allow information storage, processing and communication at faster speeds, and with lower energy consumption, than is possible with conventional electronics \cite{wolf_spintronics:_2001, awschalom_challenges_2007, chappert_emergence_2007, fert_nobel_2008}. The principal idea of spintronics is to exploit
not only the charge of an electron, but also the spin degree of freedom associated with its intrinsic angular momentum. Despite carbon not being magnetic, graphene-based spintronics may be achievable when we consider that many of its derivative materials and nanostructures display various scenarios of magnetism \cite{yazyev:review}. Graphene has also many properties that suggest its possible use as a carrier of spin information, including weak spin-orbit and hyperfine couplings, which are the main sources of relaxation and decoherence of electron spin \cite{kane_quantum_2005, huertas-hernando_spin-orbit_2006, min_intrinsic_2006, huertas-hernando_spin-orbit-mediated_2009, trauzettel_spin_2007, yazyev_hyperfine_2008, fischer_hyperfine_2009}.

Many of the proposed graphene-based spintronic devices are underpinned by the spin-polarised edge state that is predicted to occur when a graphene sheet is cut to have the so-called \emph{zigzag} edge geometry. Particular focus has been paid to zigzag-edged graphene nanoribbons, narrow strips of graphene with parallel zigzag edges that are predicted to display opposite spin orientations. Such a system may allow a possibility of tuning its spin-transport properties \cite{Nakada:1996ribbons, Nakada:1996ribbons2, Son:halfmetallic, Rossier:zigzag, Wimmer:device, Kim:device}. For example, the prospect of triggering a half-metallic state using external electric fields in zigzag-edged nanoribbons has been suggested \cite{Son:halfmetallic}. The realisation of such a device would allow efficient electronic control of spin transport, a very useful property in spintronics and something that is difficult to achieve in other materials. Despite theoretical advances in the study of nanoribbons, and some recent experimental
results hinting at signatures of this edge magnetism \cite{tao_spatially_2011}, the difficulty in patterning the edge geometries required for these effects to be observed may prevent a wider scale exploitation. Furthermore, the spin-polarised edge state for zigzag edges is predicted to be highly dependent on the edge geometry and not particularly robust under the introduction of edge disorder \cite{Kunstmann:unstable}. These factors present major obstacles in the path of utilising the intrinsic magnetic edge states of graphene in experimentally realisable devices.

Another possibility that has been proposed for graphene-based spintronics is the exploitation of defect-driven magnetic moments that arise in graphene \cite{yazyev:graphenemagnetism2, Palacios:vacancymag, Antidots:qubits}. Magnetic moments have been predicted to form around vacancies and other defects in the graphene lattice and the possibility of engineering a ferromagnetic state in graphene from such moments has been suggested. However, such a claim would seem to be restricted by the implications of the Lieb theorem \cite{Liebtheorem}, which states that any such magnetic moments arise from a disparity between the two sublattices of graphene. Large-scale, randomised disorder would tend to minimise such a disparity and prevent the formation of a ferromagnetic state. However, recent experimental evidence suggests the possibly of engineering such a state through partial hydrogenation \cite{xie_room_2011}. The existence of such a state may then be accessible through magnetoresistance measurements \cite{soriano_
magnetoresistance_2011}.

\begin{figure}[h]
\centering
\includegraphics[width =0.65\textwidth]{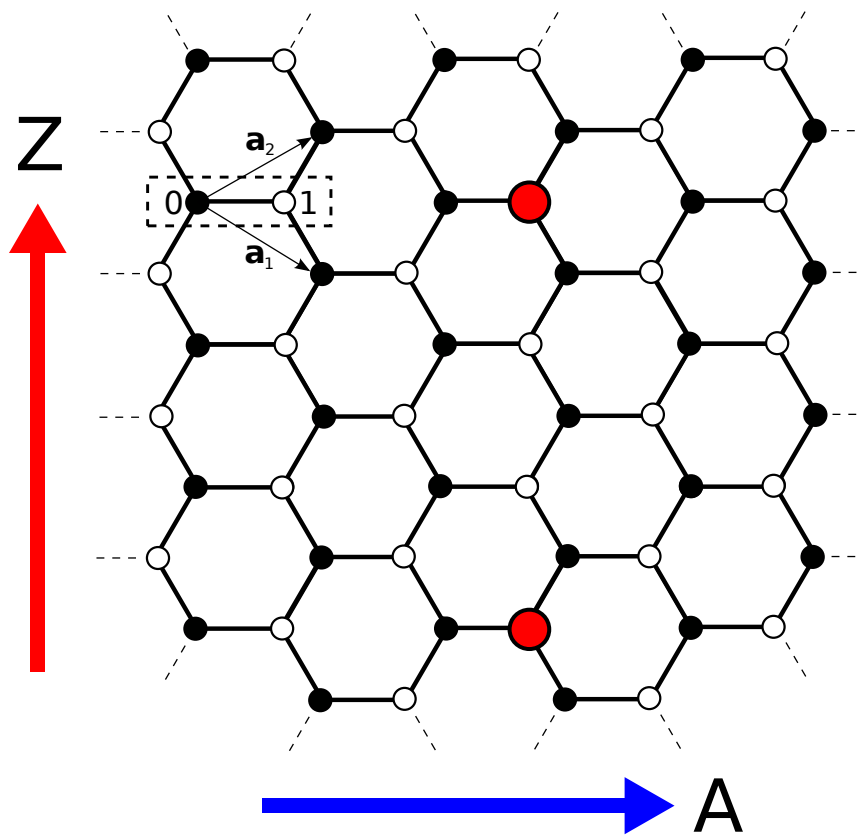}
\caption{Schematic illustration of two substitutional magnetic impurities (red circles) embedded in a graphene lattice. Also shown are the high symmetry armchair (A) and zigzag (Z) directions, and the two-atom unit cell and lattice vectors $\mathbf{a_1}$ and $\mathbf{a_2}$. Atoms from the two triangular sublattices are represented by filled and hollow symbols.}
\label{schematic}
\end{figure}

A third possibility for incorporating graphene in spintronic devices, and one that we shall focus on in this review, is through the doping of graphene with magnetic impurity atoms, as shown in Figure~\ref{schematic}. This approach allows graphene to act as host to an ensemble of transition metal atoms, or indeed more generic magnetic species, and mediate the interactions between them, potentially allowing for long ranged ordering or the transfer of spin information between impurities. In order to predict the magnetic properties of such systems, it is essential to understand both how a single impurity connects to the graphene lattice and also the nature of the interactions between multiple impurities. To address the former, a large number of studies have investigated the hybridisation and resultant electronic, magnetic and magnetotransport properties of graphene systems with different magnetic species embedded in them \cite{arkady_embedding_2009, santos_first-principles_2010, zhang_electrically_2012, PhysRevX.
1.021001, Hu12:5d}. However, the interaction between the localised moments formed at the impurity sites is predicted to be largely independent of the magnetic impurity species chosen and instead depend strongly on the electronic structure of the host material. This is because the indirect exchange coupling (IEC) \cite{Parkin:IEC, Parkin:IEC2, Edwards1, Edwards2}, often referred to as the Ruderman--Kittel--Kasuya--Yosida (RKKY) interaction \cite{RKKY:RK, RKKY:K, RKKY:Y,RKKY:Bruno1, RKKY:Bruno2}, between magnetic impurities in a graphene system is mediated by the conduction electrons of the graphene host. Throughout this work we will use the terms ``IEC'' and ``RKKY'' interchangeably to refer to a general conduction-electron mediated interaction between magnetic objects, unless specifically referring to the RKKY approximation to the interaction. A long-ranged interaction of this type allows impurity moments on graphene to feel each other's presence and respond to magnetic perturbations or excitations at other
impurity sites. The usual manifestation of this interaction is through the lowering of the total energy of the system when the moments adopt certain favourable alignments. In this way the interaction can dictate any magnetic ordering that arises between the moments. These types of interactions have been investigated previously in multilayer devices, leading to the discovery of the Giant Magnetoresistance effect \cite{Fert:GMR, Grunberg:GMR}, and more recently in materials closely related to graphene, namely carbon nanotubes~\cite{AntonioDavidIEC, David:IEC, DavidSpinValve}.

To exploit the interaction between impurities, it is important to understand its behaviour as the separation between them is varied. Previous studies suggest that this type of interaction should oscillate and decay as the impurity separation is increased \cite{Parkin:IEC, Parkin:IEC2, Edwards1, Edwards2, RKKY:RK, RKKY:K, RKKY:Y,RKKY:Bruno1, RKKY:Bruno2}. Furthermore, the oscillation period is determined by the Fermi surface of the host material, whereas the decay rate depends on the dimensionality of the host and magnetic objects. For impurity atoms in a one-dimensional or two-dimensional host medium, the interaction should be expected to decay as $\frac{1}{D}$ or $\frac{1}{D^2}$ respectively, where $D$ is the distance between the magnetic impurities. The decay rate in one-dimensional systems is borne out by investigations in metallic carbon nanotubes where the predicted long range decay is found theoretically for substitutional magnetic impurities \cite{AntonioDavidIEC}. However, unique features arising
from the peculiar electronic structure of the underlying graphene lattice were also reported. The sign of the interaction, determining the preferred alignment of the moments on two impurities, was found not to oscillate as a function of distance as seen in other materials, but rather to depend only on the connections of the impurities to the two triangular sublattices composing graphene. These sublattices are represented by filled and hollow circles in Figure \ref{schematic}. Furthermore, for certain impurity configurations, the interaction was found to decay much faster than expected \cite{David:IEC, David:PhD, DavidSpinValve}.

Similar features to those reported for carbon nanotubes have now been predicted in graphene. Due to the interest in carbon-based spintronics, a large number of studies in recent years have investigated indirect exchange interactions between localised moments in graphene. Sublattice-dependent signs and faster than expected decay rates have been seen when a variety of methods are employed, ranging from analytic or numerical treatments of the interaction using tight-binding or Dirac Hamiltonians to more intensive total energy calculations within a density functional theory (DFT) framework. In this review the details of the various approaches will be discussed in Section \ref{sec_methods} and the consensus behaviour for simple impurities that has emerged from the majority of the studies analysed in Section \ref{Results_sub}. The results for impurities beyond the simplest substitutional case and for considerations often neglected by the basic RKKY approximation are discussed in Section \ref{Results_complicated}.
Finally, in Section \ref{Results_manipulation} we consider how the interaction between magnetic impurities can be manipulated in order to extend the interaction range or otherwise alter its behaviour.

\section{Methods}

\label{sec_methods}
\begin{figure}[h]

\centering
\includegraphics[width=0.7\textwidth]{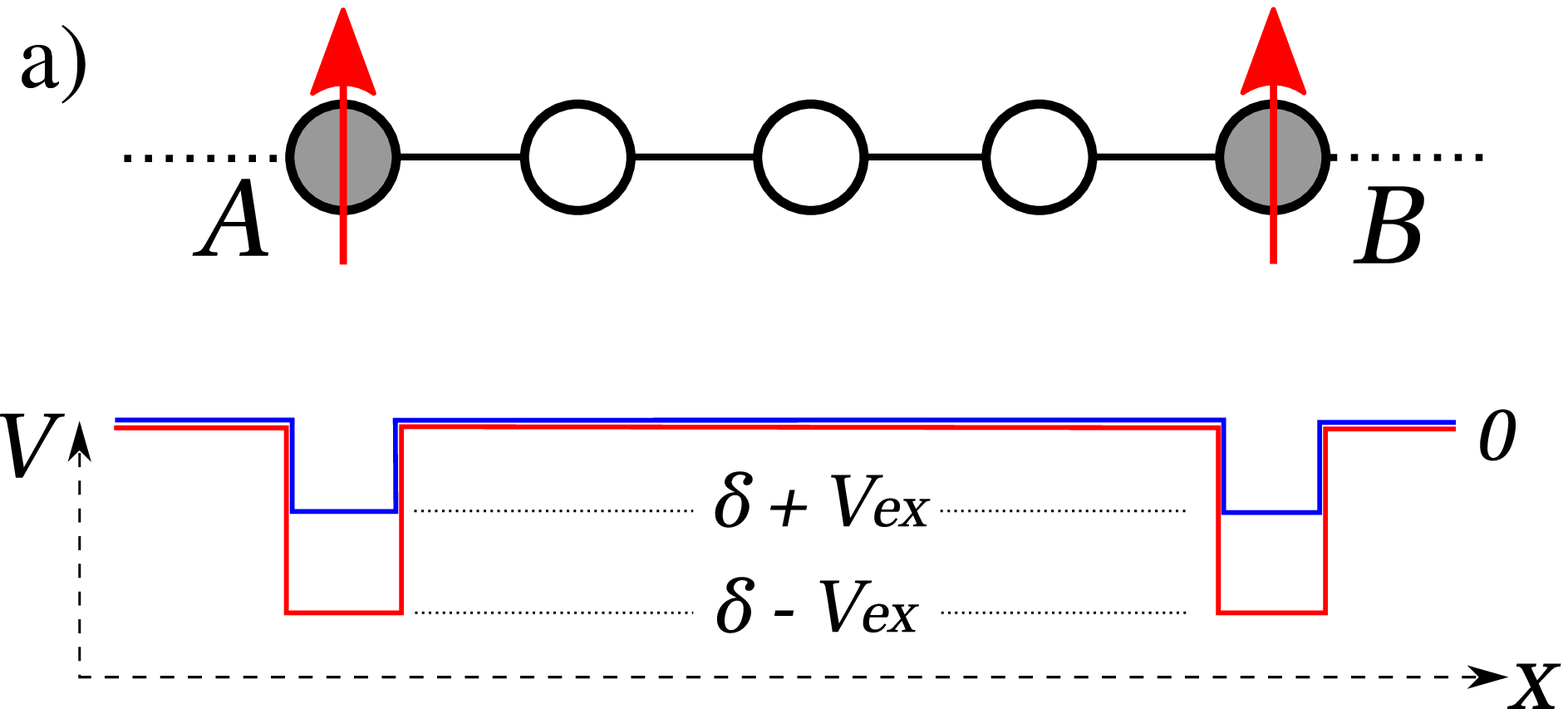}

\vspace{0.25cm}

\centering
\includegraphics[width=0.7\textwidth]{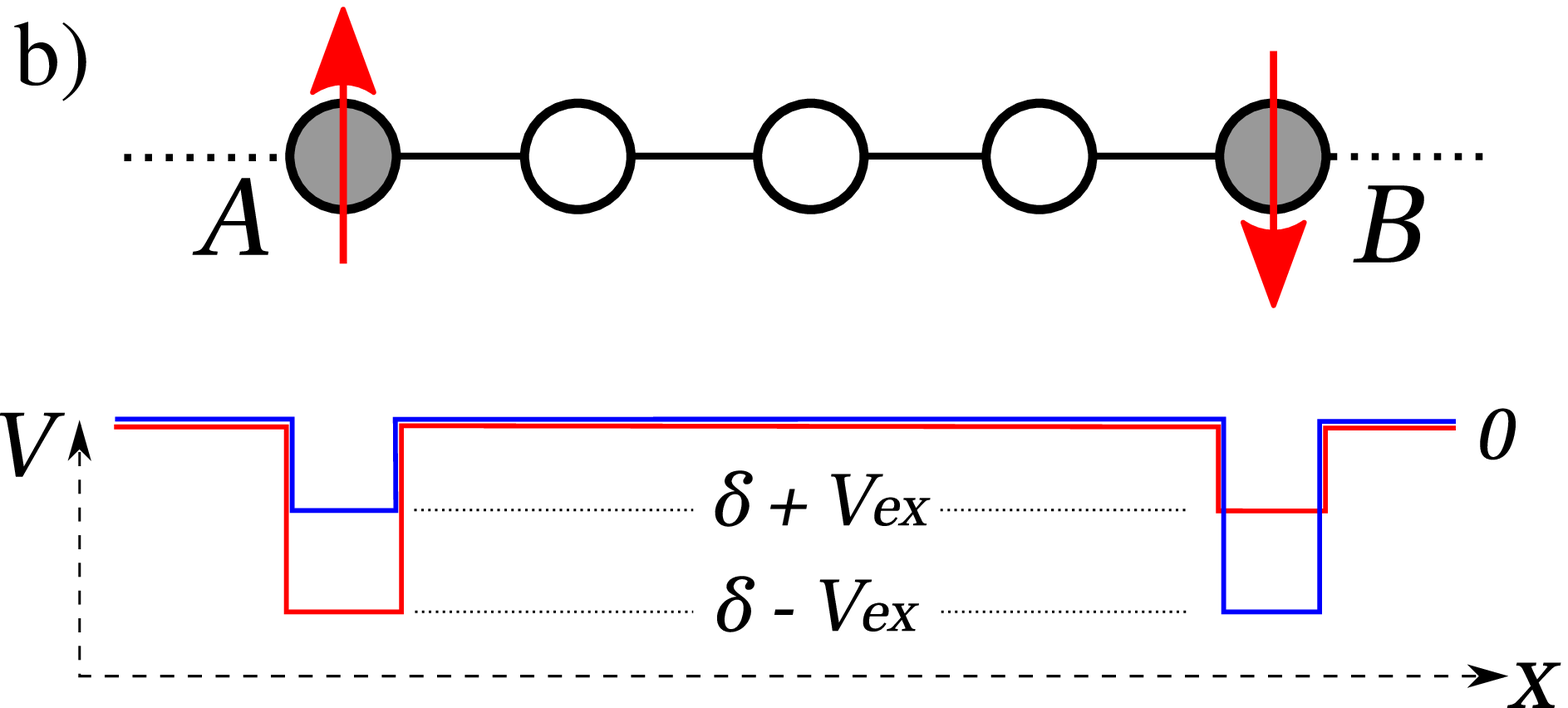}

\caption{Schematic representation of the spin-split potentials at the magnetic sites for the parallel (\textbf{a}) and antiparallel (\textbf{b}) alignments of the magnetic moments. The red (blue) curves show the potential experienced by up-spin (down-spin) electrons as a function of position. For simplicity the host medium is represented by a one-dimensional linear chain.}

\label{fig_IEC_schematic}
\end{figure}

In this section we will outline several of the general methodologies employed to calculate the indirect exchange coupling in graphene. We begin by considering an energy difference calculation between the parallel and antiparallel orientations of the localised moments on two magnetic impurities located at sites $A$ and $B$ in the graphene lattice. These configurations are represented schematically in Figure~\ref{fig_IEC_schematic}. Employing the Lloyd formula \cite{lloyd} to calculate the total energy difference between these two spin configurations we arrive at the Quantum Well method result proposed by Edwards \emph{et al.} \cite{Edwards1, Edwards2}. We shall briefly derive this result in a completely generic form in terms of Green functions without stating the form of the Hamiltonian chosen to describe the electronic structure of the graphene host or the magnetic impurities. From this result we identify the common approximations often made in the literature and demonstrate how the commonly used RKKY
expression can be reached.

From these general results we turn our attention to the specific case of graphene. Two common Hamiltonians used to describe the electronic structure, namely the tight-binding and linearised Dirac formalisms, are introduced and compared. The various approaches, numerical and analytical, to calculating the coupling within these formalisms are discussed. Finally, the merits and disadvantages of these methods, and those of more complete \emph{ab initio} treatments, are given.

\vspace {-12pt}

\subsection{Calculating Exchange Interactions}
\label{Methods_general}

The IEC between a pair of magnetic objects is usually defined as the energy difference between the parallel and antiparallel alignments of the localised magnetic moments on the objects. We distinguish the indirect interaction by assuming that it is mediated entirely by the conduction electrons of the host material, in contrast to the direct exchange which depends on the overlap of the impurity electron orbitals and which decays very abruptly. To begin, we consider an initial configuration of two magnetic impurities at sites $A$ and $B$, separated by a distance $D$, in the host medium. We introduce a Green function, $\hat{g}$, that describes the electronic structure of this system. For the moment we will keep the form of this Green function, and its associated Hamiltonian, as general as possible. We assume that it describes in sufficient detail electron propagation in the host material and also spin-dependent potentials localised at the magnetic impurities, which reflect the fact that electrons in the system
are subject to a different potential in the magnetic regions than elsewhere, as shown schematically in Figure~ \ref{fig_IEC_schematic}. The resultant potential profiles seen by ``up'' and ``down'' spin electrons are the result of solving the Hubbard Hamiltonian and are illustrated by red and blue lines respectively. The allowed energies of the system are quantised and furthermore, changing the separation between the magnetic objects shifts the allowed energies relative to the Fermi energy of the system giving rise to an oscillatory coupling \cite{quantisation_footnote}. The coupling can be calculated by summing over all the energy levels below the Fermi energy and taking the difference between the cases with spin potentials corresponding to moments aligned parallel or anti-parallel. In this manner, the calculation of the IEC reduces to the calculation of an energy difference between two distinct configurations. Such a calculation can be simplified using the Lloyd formula \cite{lloyd}, which allows the
calculation of the energy difference directly without the need to calculate the total energy of either configuration.

The Lloyd formula expression to calculate the total change in the energy of a system due to a perturbation is given by
\begin{equation}
\Delta E \, (E_F) = \frac{1}{\pi} \:\mathrm{Im} \: \int \: \mathrm{d} E \: f(E) \; \ln \left( \det (\hat{I} - \hat{g}(E) \hat{V}) \right) \
\label{lloyd}
\end{equation}
where $\hat{g}$ as above is the Green function describing the unperturbed system, $\hat{V}$ is the applied perturbation potential and $f(E)$ is the Fermi function. The unperturbed system consists of two moments embedded at sites $A$ and $B$, and aligned parallel along the z-direction so that the angle between them, $\theta = 0$. This setup is shown schematically in the panel a) of Figure \ref{fig_IEC_schematic}. For simplicity we consider one magnetic orbital on the impurity sites, but the calculation can be easily generalised to a more realistic magnetic impurity. We associate a bandcentre ($\delta$) and exchange splitting ($V_{ex}$) with this orbital as illustrated in Figure \ref{fig_IEC_schematic}.

We now introduce a spin perturbation which rotates the magnetic moment at $B$ by an angle $\theta$ with respect to that at $A$. This perturbation is given by
\begin{equation}
\begin{aligned}
V (\theta) & = - V_{ex} \, \left[ (\cos \theta - 1) \, \hat{\sigma}_z + \sin \theta \, \hat{\sigma}_x \right] \\
& = - V_{ex} \; \left[ \begin{array}{cc}
\cos \theta - 1 & \sin \theta \\
\sin \theta & 1 - \cos \theta
\end{array}
\right] \
\end{aligned}
\end{equation}
where $\hat{\sigma}_z$ and $\hat{\sigma}_x$ are the relevant Pauli matrices. Since the magnetic moments break the spin degeneracy of the electrons, the Green functions must be written in terms of their up- and down-spin components. In the initial collinear configuration there is no mixing between the spin bands and $g_{BB}$ is diagonal in spin space. It can be shown that
\begin{equation}
\det \left[ \hat{I} - \hat{g}_{BB}(E) \, \hat{V (\theta)} \right] = 1 - 2 V_{ex}^2 \; g_{BA}^{\uparrow} \; g_{AB}^{\downarrow} \; (\cos \theta -1) \
\label{IECdet}
\end{equation}
where $g_{AB}^{\uparrow}$ ($g_{AB}^{\downarrow}$) are the off-diagonal elements of the up (down) spin Green function connecting sites $A$ and $B$. Taking $\theta = \pi$, corresponding to an antiparallel alignment of the moments we find, for zero temperature, that
\begin{equation}
J_{BA} = - \Delta E (\theta = \pi) = - \frac{1}{\pi} \:\mathrm{Im} \: \int_{-\infty}^{E_F} \: \mathrm{d} E \: \ln \left( 1 + 4 \; V_{ex}^2 \; g_{BA}^{\uparrow} (E) \; g_{AB}^{\downarrow} (E) \right) \
\label{staticJ}
\end{equation}
where we have chosen the sign convention where a negative value of the coupling, $J$, corresponds to a preferential parallel alignment of the moments and a positive value to a preferential antiparallel alignment. When performing the integral in Equation \eqref{staticJ} numerically we can rewrite the integral over the imaginary axis where the integrand tends to be smoother and easier to integrate. With an imaginary axis integration, the expression for the coupling becomes
\begin{equation}
J_{BA} = \frac{1}{\pi} \: \int_{\eta}^{\infty} \: \mathrm{d} y \: \ln \left| 1 + 4 \; V_{ex}^2 \; g_{BA}^{\uparrow} (E_F + iy) \; g_{AB}^{\downarrow} (E_F + iy) \right| \
\label{staticJimag}
\end{equation}
We note that in these expressions the distance dependence is entirely contained within the product of off-diagonal Greens function elements and the behaviour of these quantities will dictate the behaviour of the interaction as the separation between the moments is varied.

The expression for the coupling given in Equation \eqref{staticJ}, with the log term in the integrand, is exact. That is, it gives the exact energy difference between the FM and AFM configurations of the system described by the Green functions used. In the next section we will discuss how the relevant Green functions will be calculated, but first we focus on a common perturbative approach to calculating the coupling. To proceed analytically it is useful to note that Equation \eqref{staticJ} can be written as a perturbation expansion in powers of the exchange splitting $V_{ex}$ and when expressed to leading order in $V_{ex}$ gives
\begin{equation}
J_{BA} = - \frac{4 \, V_{ex}^2}{\pi} \, \int \, \mathrm{d} E \, f(E) \, \mathrm{Im} \left[\, g_{BA} (E) \, g_{AB} (E) \, \right] \
\label{static_J_rkky}
\end{equation}
where $g_{BA} (E)$ is the spin-independent Green Function describing electron propagation in the pristine host material \cite{Castro:Coupling}.

This expression for the coupling is equivalent to the Ruderman--Kittel--Kasuya--Yosida (RKKY) approach, initially developed to describe the coupling mechanism of nuclear magnetic moments \cite{RKKY:RK}, then expanded to describe a wider range of indirect coupling phenomena \cite{RKKY:K, RKKY:Y} and generalised to provide a model capable of reproducing experimental observations \cite{RKKY:Bruno1, RKKY:Bruno2}. This treats the coupling as a consequence of the spin polarisation of the conduction electrons of the host by the magnetic objects. Under this approach, the coupling is written as an effective direct coupling, $J_{BA} \; \mathbf{s}_B \cdot \mathbf{s}_A$, between the two moments, where the coupling strength is given by
\begin{equation}
J_{BA} = \left( \frac{\lambda^2 \, \hbar^2}{4}\right) \; \chi^0_{BA} \
\label{rkky_chi}
\end{equation}
where $\lambda$ is an adjustable parameter representing the magnitude of the coupling between localised spins and conduction electrons and $\chi$ is the static magnetic susceptibility which relates the response of the host magnetisation to a static magnetic field. Writing this quantity in terms of Green functions,
\begin{equation}
\chi^0_{BA} = -\frac{2}{\pi} \; \int \; \mathrm{d} E \; f(E) \; \mathrm{Im} \left[\, g_{BA} (E) \; g_{AB} (E) \, \right] \
\label{static_susceptibility}
\end{equation}
and the equivalence of Equations \eqref{static_J_rkky} and \eqref{rkky_chi} becomes clear.

Thus the RKKY approach can be considered a second-order perturbational approximation to the IEC. The approaches are in general equivalent, except when $V_{ex} >> 1$ or when the host material possesses certain symmetries \cite{Castro:Coupling, Mauro:RKKY1, Mauro:RKKY2, Tung199696}. The RKKY approximation generally provides a good description at large separations when the coupling is quite small. An important contrast between the Lloyd formula approach and the RKKY approach is the choice of Green function used. The former approach avails of the spin dependent Green functions in the ferromagnetic configuration, $g_{BA}^{\uparrow}$ and $g_{AB}^{\downarrow}$ whereas the latter approach generally uses their pristine, spin-independent counterparts. We note that use of the pristine GFs essentially sets the value of the bandcentre, $\delta$, to zero. An intermediate approach can be taken by using the spin-dependent Green functions with an RKKY-like expansion of the logarithm in the Lloyd expression. The general form
expected for the coupling is
\begin{equation}
J ( D ) \sim \frac{\cos (2 \mathbf{k_F} \cdot \mathbf{D} ) } {D^{\alpha}}
\end{equation}
where $\mathbf{k_F}$ is the Fermi wavevector in the conducting host and the exponent $\alpha$ determining the rate of decay with separation depends on the dimensionality of the system.

Most of the analytic investigations into the IEC in graphene make use of the RKKY approximation. Although we have written it here in terms of energy dependent Green functions, it is often formulated differently. A common alternative formulation is in terms of time dependent Green functions, the Fourier transforms of the GFs used here, although conflicting results have been reported depending on whether the real or imaginary time representations are used \cite{saremi:graphenerkky, bunder:rkkygraphene, kogan:rkkygraphene}. Numerical approaches to indirect exchange interaction generally proceed in one of two ways. The first uses numerical methods to calculate either perturbative or exact expressions for the coupling like those given above. Another method involves direct calculation of the energies of the individual FM and AFM configurations. This is a common approach in \emph{ab initio} calculations \cite{Lehtinen_diffusionPRL, pisani_rkkydft, santos_magnetism_2010, soriano_hGNRs, arkady_rkkydft} although it
has been employed also for tight-binding models \cite{black:graphenerkky, black-schaffer_importance_2010}. The computational cost of this type of calculation tends to limit these investigations to small or medium values of separation and it can be difficult to perform a meaningful comparison with results for the asymptotic behaviour calculated in other ways.

\subsection{Calculations in Graphene}
\label{Methods_graphene}

Now that we have introduced the general methodology to calculate indirect exchange couplings, we shift our focus to examine the specific case of interactions in graphene. The electronic structure of graphene plays a key role in determining the interaction properties and so we first outline the different methods used to calculate or approximate the band structure. We then discuss how the Green functions required for some coupling expressions can be found and compare the different approaches used to calculate the coupling.

The electronic structure of graphene near the Fermi energy is principally determined by $\pi$ bonds formed from the out of plane $2 p_z$ orbitals of carbon atoms situated on the vertices of a hexagonal, or honeycomb, lattice. The remaining carbon orbitals hybridize in plane to form very stable $\sigma$ bonds which give graphene its extraordinary strength. Thus a single-orbital nearest-neighbour tight-binding Hamiltonian is usually more than adequate when describing the electronic properties of graphene. To write this Hamiltonian we make use of the two-atom unit cell shown in Figure \ref{schematic}. Thus every site in the lattice is denoted by a vector $\mathbf{r}$ giving the unit cell location and an index $n = 0,1$ differentiating between the two sites within the unit cell, illustrated by filled and hollow symbols in the figure. The tight-binding Hamiltonian is written
\begin{equation}
\hat{H} = \sum_{<\mathbf{r}, n, \mathbf{r}^\prime, n^\prime>} | \mathbf{r} n \rangle \, t_{\mathbf{r},\mathbf{r}^\prime}^{n, n^\prime} \, \langle \mathbf{r}^\prime n^\prime |
\label{hamiltonian}
\end{equation}
where $| \mathbf{r} n \rangle$ represents the orbital at the site given by $\{\mathbf{r}, n\}$ and $t_{\mathbf{r},\mathbf{r}^\prime}^{n, n^\prime}$ is the electronic hopping term between two such orbitals. The sum is restricted to orbitals at neighbouring sites and the non-zero hopping terms all take the value $t = - 2.7 \,\mathrm{eV}$. One consequence of only allowing nearest neighbour hoppings is that graphene is a bipartite lattice. The sets of filled and hollow symbols in Figure \ref{schematic} can be regarded as intersecting triangular sublattices that together form the honeycomb graphene lattice. Introducing hopping terms beyond the first nearest neighbour break the bipartiteness of graphene. However, since these terms are much smaller than the nearest neighbour hoppings, graphene behaves effectively as a bipartite lattice. It is this feature of the graphene lattice, combined with the half filling of its $\pi$ orbitals, that determine many of the features of the RKKY interaction \cite{saremi:
graphenerkky}.

\begin{figure}[h]
\centering
\includegraphics[width =0.95\textwidth]{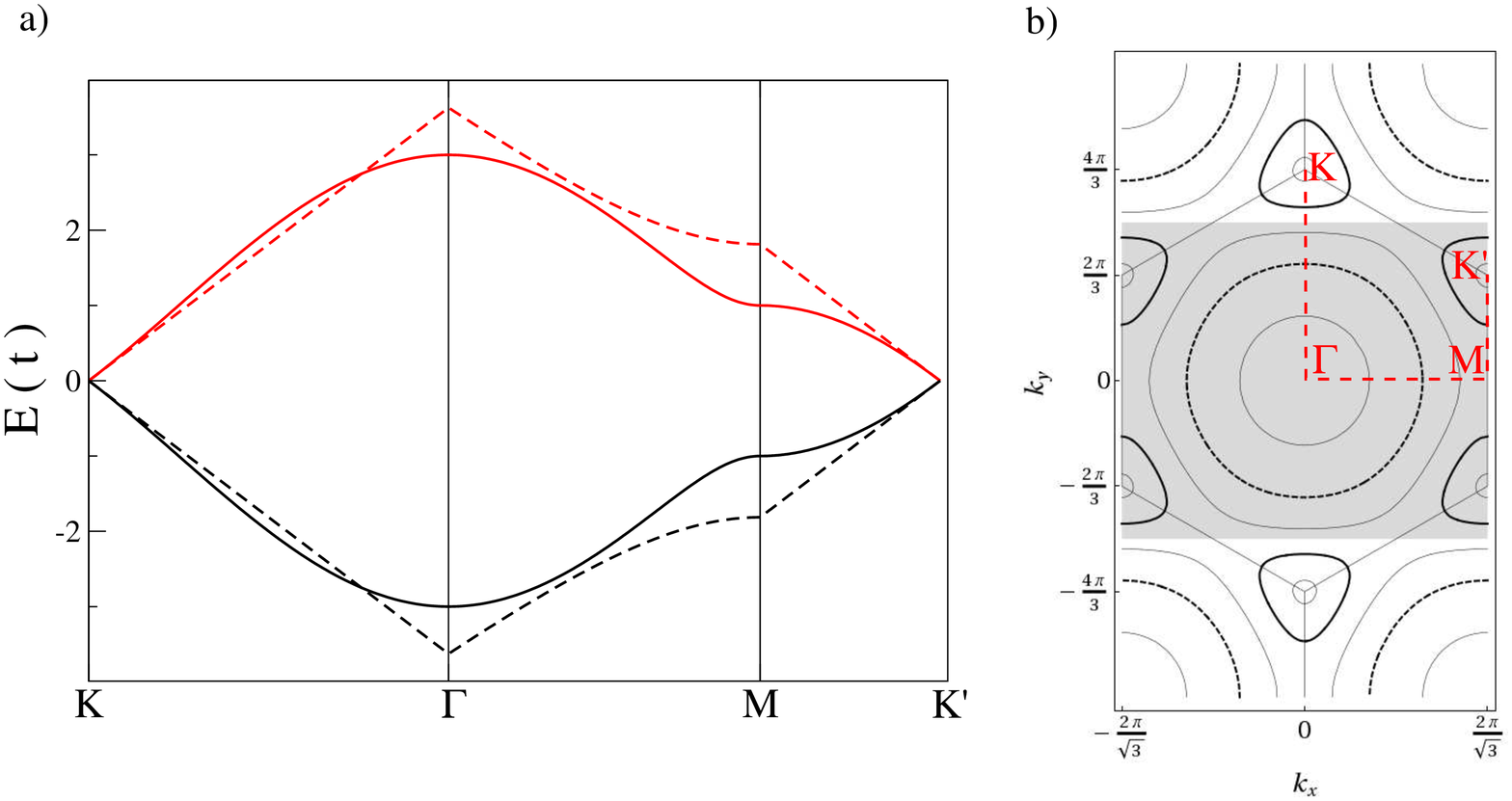}
\caption{(\textbf{a}) shows the band structure of graphene calculated using both the tight-binding Hamiltonian (solid lines) and the linearised Hamiltonian (dashed lines). The bands are plotted along the high symmetry points of the graphene Brillouin Zone shown by the dashed red plot in (\textbf{b}). The shaded rectangular area in (\textbf{b}) is a possible alternative to the standard hexagonal Brillouin zone. }
\label{bandsandBZ}
\end{figure}

Using Bloch theorem methods to take advantage of the lattice periodicity, the associated eigenvalues and eigenvectors of Equation \eqref{hamiltonian} are found to be
\begin{equation}
\epsilon_\pm (\mathbf{k}) = \pm t \, |f(\mathbf{k})|
\label{graphene_bands}
\end{equation}
and
\begin{equation}
| \mathbf{k}, \pm \rangle = \frac{1}{\sqrt{2N}} \, \sum_\mathbf{r} \, e^{-i \mathbf{k} \cdot \mathbf{r}} \, \left( \begin{array}{c} 1 \\ \pm e^{-i \phi(\mathbf{k})} \end{array} \right) \
\label{graphene_states} \vspace {6pt}
\end{equation}
where $e^{-i \phi(\mathbf{k})} = \frac{f^*(\mathbf{k})}{|f(\mathbf{k})|}$ and $f(\mathbf{k}) = e^{-i \frac{k_x a}{\sqrt{3}} } \, + \, 2 \cos(\frac{k_y a}{2}) \, e^{ i \frac{k_x a}{2 \sqrt{3}} }$. The bands given by Equation \eqref{graphene_bands} are shown plotted along the high symmetry points of the graphene Brillouin zone by the solid lines in Figure~\ref{bandsandBZ}a.

We note that since the $p_z$ orbitals in carbon contain one electron each, then the band is half full and the Fermi energy of the undoped system is $E_F = 0$. This is exactly the point where the two bands given by $\epsilon_-$ and $\epsilon_+$ touch and the DOS vanishes. Graphene is thus a zero-bandgap semiconductor, or a \emph{semi-metal}. The resultant Fermi surface consists of six discrete points, only two of which are unique, lying at the vertices, or $K$ points, of the hexagonal first Brillouin Zone (BZ) illustrated in Figure \ref{bandsandBZ}. Many of the interesting properties belonging to graphene arise due to the shape of the bands near the Fermi energy. Unlike in conventional semiconductors where the bands are parabolic, the band structure of graphene is linear near $E_F$. This results in electrons or holes near the Dirac points having zero effective mass and behaving like relativistic particles which can be described using the Dirac equation from Quantum Electrodynamics. It is worth investigating
briefly how graphene electrons are approximated in the linear regime and the range of energy values over which the approximation is valid.

The region of the graphene spectrum around $E_F = 0$ can be approximated linearly to simplify the calculation of physical properties. For a point in reciprocal space near the Dirac point, $\mathbf{k} = \mathbf{K} + \mathbf{\delta k}$, we~find
\begin{equation}
\epsilon_\pm \; (| \mathbf{\delta k} |)= \pm \frac{\sqrt{3} \, a \, t}{2} \; | \mathbf{\delta k} | \ = \pm \hbar \, v_F \, | \mathbf{\delta k} | \ \vspace {6pt}
\end{equation}
where the Fermi velocity of graphene, $v_F = \frac{1}{\hbar} \frac{\mathrm{d} E} {\mathrm{d} k} \sim 10^6 \mathrm{\,m\,s^{-1}}$. In Figure \ref{bandsandBZ}a we compare the linear approximation of the band structure (dashed lines) with that calculated previously (solid lines). It is seen to be a very good approximation in the region surrounding $E = 0$, but loses accuracy quickly outside this regime.

To calculate the coupling using the expressions in the previous section, we need the Green function connecting two sites on the pristine graphene lattice. There are a number of different ways to calculate this quantity and use it to calculate the coupling, but for illustration we will focus on a real space, energy dependent Green function calculated from the dispersion relation and eigenvectors of the non-linearised tight-binding Hamiltonian. The Green function we require is given by an expression of the form
\begin{equation}
\begin{split}
g_{jl}^{n_j, n_l} & = \langle \mathbf{r}_j, n_j \, |\, \hat{g} (E) \, | \, \mathbf{r}_l , n_l \rangle \\
& = \frac{a}{2 \pi} \frac{a \sqrt{3}}{4 \pi} \, \int \mathrm{d} k_y \, \int \mathrm{d} k_x \, \frac{ N (E) \, e^{i \mathbf{k} \cdot (\mathbf{r_l - r_j}) } } {E^2 -t^2 \,|f(\mathbf{k}) |^2 } \
\end{split}
\label{gf_int_1}
\end{equation}
where
\begin{equation*}
\begin{split}
N (E) = \quad & E \qquad \mathrm{if \qquad } n_j = n_l\\
& t f(\mathbf{k}) \,\,\qquad n_j=1, n_l=2 \\
& t f^*(\mathbf{k}) \qquad n_j=2, n_l=1 \
\end{split}
\end{equation*}
and the reciprocal space integral is over the first BZ, where for convenience, we can define a rectangular BZ consisting of segments drawn from multiple neighbouring BZs whose area equals that of the hexagonal BZ. A possible rectangular BZ is shown by the shaded area in Figure \ref{bandsandBZ}b. We note that the Green function expression takes a different form for sites on the same or on opposite sublattices. The most obvious, if not most efficient, way to proceed from here is a brute-force two-dimensional numerical integration. However, we have shown previously \cite{me:grapheneGF} that it is possible to perform one of the integrals completely analytically, greatly reducing the numerical cost of the Green function calculation. Furthermore, for separations greater than a few lattice spacings in the high symmetry zigzag and armchair directions it is possible to approximate the remaining integration to a high degree of accuracy using the Stationary Phase Approximation (SPA), to yield a closed form expression
for the Green function
\begin{equation}
{\cal G}_{D}(E) = {{\cal A}(E) \, e^{i \mathcal{Q} (E) D} \over \sqrt{D}} \
\label{concise}
\end{equation}
where ${\cal A}(E)$ is an energy-dependent coefficient and $\mathcal{Q} (E)$ can be identified with the Fermi wavevector in the direction of separation. The full analytical expressions for these quantities for armchair and zigzag separations can be found in \cite{me:grapheneGF}. Once calculated using any of these methods, the Green functions can be substituted directly into the RKKY integration in Equation \eqref{static_J_rkky}. Alternatively, they can be used to calculate the spin-dependent Green functions required for the non-perturbative total energy expression in Equation \eqref{staticJ}, where the Dyson equation is used to add a spin-dependent potential at the magnetic~sites.

At this stage, it is worth mentioning some of the alternative methods used to calculate the graphene Green function for RKKY calculations in the literature. Many of the earlier, analytic studies employ the linear approximation to the graphene band structure \cite{Vozmediano:2005, PhysRevLett.97.226801, saremi:graphenerkky, brey:graphenerkky, bunder:rkkygraphene} and may also use time-dependent Green functions in lieu of the energy dependent form shown here \cite{saremi:graphenerkky, bunder:rkkygraphene, kogan:rkkygraphene}. There is some discrepancy between early results for both the sign and decay rate of the interaction (e.g., \cite{PhysRevLett.97.226801, saremi:graphenerkky, bunder:rkkygraphene}), but many of the key features to be discussed in the next section are present in \cite{saremi:graphenerkky} and later confirmed by other studies. Numerical investigations using Green function techniques have tended to use an energy dependent form, but calculated in a variety of ways \cite{me:grapheneGF, sherafati:
graphenerkky, disorderedRKKY}. Sherafati and Satpathy \cite{sherafati:graphenerkky} employ both full numerical integration of the Green function integral in Equation \eqref{gf_int_1} and the recursive Horiguchi method, which relates the graphene Green function to those of the triangular lattice, which in turn are expressed in terms of elliptic integrals \cite{horiguchi}. Although more efficient than the full integration, the Horiguchi method was only stable for small values of separation \cite{sherafati:graphenerkky}. In the same work, an analytic treatment in the linear dispersion regime using energy dependent GFs confirmed, with some refinement, many of the features predicted previously using time dependent GFs. In a study of RKKY interactions in disordered graphene sheets, Lee \emph{et al.} \cite{disorderedRKKY} used the Kernel Polynomial Method \cite{kernelpm} to express the Green functions in terms of Chebyshev polynomials with recursively calculated coefficients. Working outside the Green function
formalism the interaction can be calculated by determining the total energies of the FM and AFM alignments of the moments. Such an approach can use tight-binding models similar to those presented here \cite{black:graphenerkky, black-schaffer_importance_2010} or more complete \emph{ab initio} calculations \cite{Lehtinen_diffusionPRL, pisani_rkkydft, santos_magnetism_2010, soriano_hGNRs, arkady_rkkydft}. Such methods often rely on periodic boundary conditions and relatively small unit cells, leading to problems like interference between moments in neighbouring cells \cite{rapidcomm:emergence} and difficulty in calculating the interaction for separations greater than a few multiples of the lattice parameter.


\section{Interaction between Simple Magnetic Impurities}
\label{Results_sub}

In this section we consider localised magnetic moments that are associated with a single lattice site in graphene. These can be thought of as substitutional magnetic impurities replacing a single carbon atom in the lattice, or as moments induced at a lattice site by the presence of an adsorbate. This type of impurity is the most commonly considered in the literature and the exchange interactions between pairs of these impurities have been investigated using many of the methods discussed in the previous sections. The expressions for the couplings given in Equations \eqref{staticJ} and \eqref{static_J_rkky} are easily applied to moments of this kind as the Green functions required are either those for pristine graphene with a simple \linebreak spin-dependent perturbation at the magnetic sites, or without this perturbation if the RKKY expression is used. Throughout this section we will consider two magnetic impurities occupying sites $A$ and $B$ in the graphene lattice, separated by a distance $D$, with an
indirect exchange interaction $J_{AB}$.

In the earliest studies, the RKKY interaction in graphene was often considered in conjunction with Friedel Oscillations (FO). The latter involve distance dependent charge fluctuations relative to an impurity site and there is a large degree of overlap in the physics involved. Using an analytic Dirac linearisation approach to study FO, Cheianov and Fal$'$ko predicted that charge density FO would decay as $\delta \rho \sim D^{-3}$, but the RKKY interaction as $J \sim D^{-2}$ \cite{PhysRevLett.97.226801}. Wunsch \emph{et al.}, meanwhile, found the magnitude of induced moments to decay as $D^{-3}$ away from a magnetic impurity and predicted the RKKY interaction to behave likewise \cite{wunsch2006}. Before these reports, an early study by Vozmediano~\emph{et~al.} investigating possible ferromagnetism in graphene layers arising from localised moments near lattice distortions had examined the RKKY interaction between moments surrounding elongated cracks in graphene \cite{Vozmediano:2005}. Although more complicated
structures than considered elsewhere, a decay rate of $D^{-3}$ for a non-oscillatory, always ferromagnetic interaction was reported. An important paper by Saremi clarified many of the features of the interaction \cite{saremi:graphenerkky}. Firstly, a general proof was given regarding the sign of the RKKY interaction in a half-filled bipartite lattice. This showed that, due to electron-hole symmetry in such systems, the coupling was ferromagnetic between moments on the same sublattice and antiferromagnetic between moments on opposite sublattices. In the same work, analytic calculations were performed in the linear dispersion regime to determine the functional form of the RKKY interaction. These calculations included the explicit use of a cut-off function to prevent divergences arising from high energy contributions. There has since been some debate about the form, or even the necessity, of the cut-off function and of the nature in which the time-dependent Green function integrations are performed \cite{saremi:
graphenerkky, bunder:rkkygraphene, black:graphenerkky, sherafati:graphenerkky, me:grapheneGF, kogan:rkkygraphene, kogan2012all}. However, many of the principal features of the interaction reported by Saremi have been verified by other studies, including a contemporaneous paper by Brey \emph{et al.} \cite{brey:graphenerkky}.

\begin{enumerate}
\item FM (AFM) interaction between moments on the same (opposite) sublattice(s), with the magnitude of the AFM interaction three times greater; \vspace {-6pt}
\item A decay rate of $J_{AB} \sim D^{-3}$;\vspace {-6pt}
\item An oscillatory term of the form $1 + \cos (2 \mathcal{Q} (E_F) D)$, where $\mathcal{Q} (E_F)$ is the component of the Fermi wavevector parallel to the separation direction.
\end{enumerate}

Bunder and Lin suggested that differences between the real time and imaginary time approaches could lead to a breaking of the theorem regarding the sign of the interaction \cite{bunder:rkkygraphene}. In particular they report sign changing oscillations for the interaction in zigzag-edged graphene nanoribbons. However, fully-numerical calculations of the coupling by Black-Schaffer do not find such sign changes, which the author attributes to the perturbative nature of the previous calculations \cite{black:graphenerkky}. In the same work, the main predictions made by Saremi are verified using a numerical exact diagonalisation method to find the energy difference between the FM and AFM orientations. A minor refinement is the introduction of a $\pi$ phase shift in the cosine term for moments on opposite sublattices. In fact, Sherafati and Satpathy showed that this phase shift is not a constant but depends on the direction of separation, taking a maximum value of $\pi$ for zigzag separations and vanishing for
armchair separations \cite{sherafati:graphenerkky}. This additional phase shift term can be added to the three features proposed by Saremi to provide a complete description of the RKKY interaction between simple single-site magnetic moments in graphene.

Using the stationary phase form of the Green function in Equation \eqref{concise}, we have previously demonstrated how the principal features of the interaction in the RKKY approximation can be calculated~\cite{me:grapheneGF}. The integration over energy in Equation \eqref{staticJ} can be reduced to a sum over Matsubara frequencies. The functions ${\cal B}(E, \varepsilon) = {\cal A}^2(E, \varepsilon)$ and ${\cal Q}(E, \varepsilon)$ are expanded around $E_F$ and in the low temperature limit $T \rightarrow 0$, we find
\begin{equation}
J_{BA} \sim \mathrm{Im} \,\sum_{\ell} \frac{\mathcal{J}_\ell (E_F) }{D^{\,\ell+2}} \, \exp\,(2 i \mathcal{Q}(E_F)\, D)
\label{spa_coupling}
\end{equation}
where
\begin{equation}
\qquad \mathcal{J}_\ell (E_F ) = \frac{(-1)^\ell \, V_{ex}^2 \, \mathcal{B}^{(\ell)} (E_F ) }{(2 \mathcal{Q}^\prime (E_F ))^{\,\ell+1} }
\label{coupling_coeff}\vspace {6pt}
\end{equation}
is the distance-independent coefficient for the $\ell$-th term in the power series, $\ell$ is a non-negative integer and $\mathcal{B}^{(\ell)} (E_F )$ is the $\ell$-th order energy derivative of $\mathcal{B} (E )$ evaluated at $E_F$, resulting from its Taylor expansion. In general the leading term in the series should determine the asymptotic decay rate of the coupling. For the undoped case it can be shown that the coefficient $\mathcal{B}^{(0)} (0 ) =0$, so that the $l=1$ term dominates and $J(E_F=0) \sim D^{-3}$. The exact oscillatory term predicted by Saremi is reached by forcing the real part of the Green function to vanish at $E_F=0$.

\begin{figure}[h!]
\centering
\includegraphics[width =0.7\textwidth]{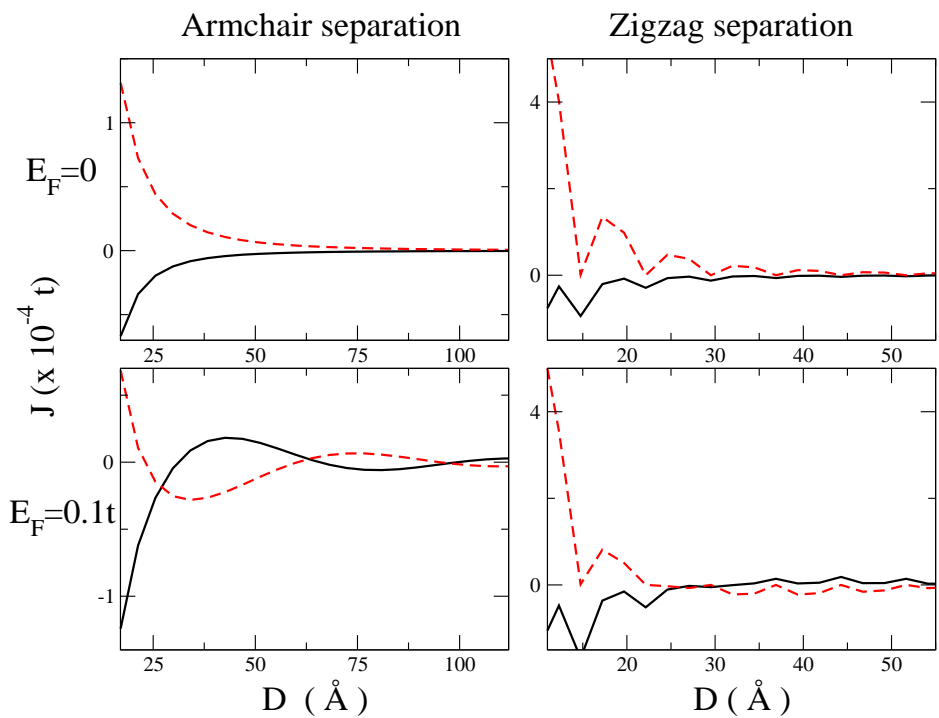}
\caption{The indirect exchange coupling between two moments separated in the armchair (direction) are shown in the left (right) panel as a function of moment separation. Results for the undoped (doped) case are shown in the top (bottom) panels. Results are shown for both the same sublattice (black, solid lines) and opposite sublattice (red, dashed lines) cases.}
\label{fig_cases}
\end{figure}

Figure \ref{fig_cases} shows a numerical calculation of the coupling, calculated using Equation \eqref{staticJimag}, for substitutional magnetic impurities in graphene. The coupling is plotted as a function of separation for both armchair and zigzag directions, and for impurities on the same and on opposite sublattices. Results are shown for both undoped ($E_F = 0.0$) and doped ($E_F=0.1|t|$) cases. The full range of features discussed in this section are present. In particular we note the $D^{-3}$ decay rate and absence of sign-changing oscillations in the undoped cases. We note the presence of a period-3 oscillation in the zigzag direction compared to the monotonic decay for the armchair case. This results from slightly different commensurability effects in the two directions. In undoped graphene, the oscillations are determined by $\cos^2 ( \mathbf{k_F} \cdot \mathbf{D})$. The components of the Fermi wavevector in the armchair and zigzag directions are $\frac{2\pi}{\sqrt{3}a}$ and $\frac{2\pi}{3a}$
respectively. Since the spacing between unit cells is $\sqrt{3}a$ for armchair and $a$ for zigzag separations, the cosine squared term takes a constant value of $1$ in the armchair direction but cycles repeatedly through the sequence $\tfrac{1}{4}$, $\tfrac{1}{4}$, $1$ as the separation is increased in the zigzag direction. Also evident in the zigzag separation plot is the $\pi$ phase shift between same and opposite sublattice cases. When the system is doped, as in the bottom panels, a longer-ranged interaction decaying as $J \sim D^{-2}$ is seen. Long-wavelength, sign-changing oscillations are also introduced by the change in the shape of the doped Fermi surface.

\pagebreak
\section{Beyond the Simple Case}
\label{Results_complicated}

In the previous section we highlighted the features of the magnetic interaction in graphene for the simplest kind of impurities. These impurities are the kind usually investigated within the RKKY approximation. In this section we consider a number of extensions to the simple Hamiltonian describing the system and examine the effects they have on the coupling. We consider first a more general or realistic description of the magnetic impurities and examine the effect of impurity parameterisation on the interaction. We show that the use of the pristine Green function in the RKKY approximation ignores these effects which can in principle overturn some of the results noted in the previous section. Then we move to the case when the impurity connects to more than one atom in the graphene lattice and examine the effect on the decay rate of the interaction. Finally, more detailed descriptions of the graphene electronic structure are considered. We discuss the possible effects of electron-electron interactions and spin-
orbit coupling and examine the results of some \emph{ab initio} calculations.

\subsection{Impurity Parameterisation}

Calculations using the RKKY approximation do not generally account properly for the local spin dependent potentials describing the magnetic moments. In fact, the parameterisation of the magnetic impurity is generally neglected through the usage of the pristine Green functions. In this section we consider the effect of three parameters that can be used to characterise the magnetic impurities. In addition to the magnetic moment ($m$) and band-centre shift ($\delta$), shown in Figure \ref{fig_IEC_schematic} and which can be calculated, for example, using self-consistent mean-field calculations, we also consider the hopping potential between the lattice carbon sites and the impurity site ($t^\prime$) which should differ from the \linebreak carbon-carbon hopping. This approach is similar to the Anderson model describing localised magnetic impurity states in metals \cite{anderson_localized_1961}. These parameters will vary between the different magnetic species that can be chosen as the embedded impurities.

\begin{figure}[h]
\centering
\includegraphics[width=0.65\textwidth]{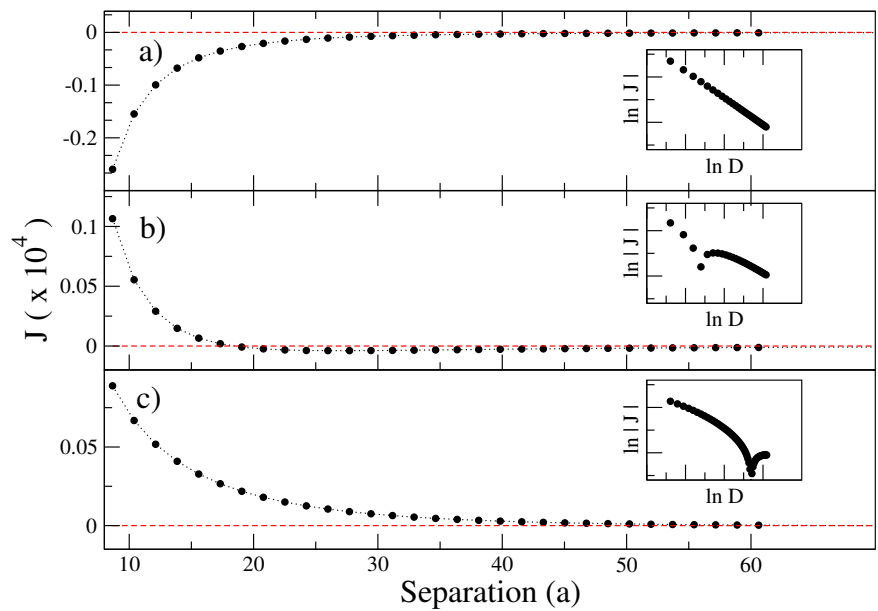}
\caption{The magnetic coupling between two magnetic sites on the same sublattice as a function of their separation, $D$, along the armchair direction for three different impurity parameterisations: (\textbf{a}) ($m=0.6$, $\delta=0.0$, $t^\prime = t$); (\textbf{b}) ($m=0.6$, $\delta=5.0t$, $t^\prime = 0.8t$) and (\textbf{c}) ($m=0.6$, $\delta=8.0t$, $t^\prime = 0.6t$). The insets show log-log plots where a sign change in the coupling is evident from a dip feature. }
\label{fig_signchange1}
\end{figure}

In Figure \ref{fig_signchange1} we plot the coupling between substitutional impurities on the same sublattice, calculated using the Equation \eqref{staticJ} and exact Green functions, as a function of separation along the armchair direction for three different parameter sets $(m, \delta, t^\prime)$. The first of these $(0.6, 0.0, t)$ closely replicates the results of the RKKY approach as it considers only a band-splitting, has no band-centre shift and uses the carbon-carbon hopping value. The numerical calculations in the previous section used a similar parameterisation. The middle and bottom plots use the parameters $(0.6, 5.0, 0.8t)$ and $(0.6, 8.0, 0.6t)$ respectively. In these plots we see the formation of an unusual feature not predicted by the RKKY approximation. For quite a large range of distances we note a preferential \emph{antiferromagnetic} alignment between the moments before the sign flips and the standard ferromagnetic coupling with a $1/D^3$ decay is recovered. A similar sign-changing
behaviour has been reported in nanotubes \cite{DavidSpinValve} and also by \emph{ab initio} calculations attempting to probe the interaction in graphene \cite{arkady_rkkydft}. It is worth examining further how this feature depends on the parameterisation of the magnetic moments.

\begin{figure}[h]\vspace {-36pt}
\centering
\includegraphics[width=0.85\textwidth]{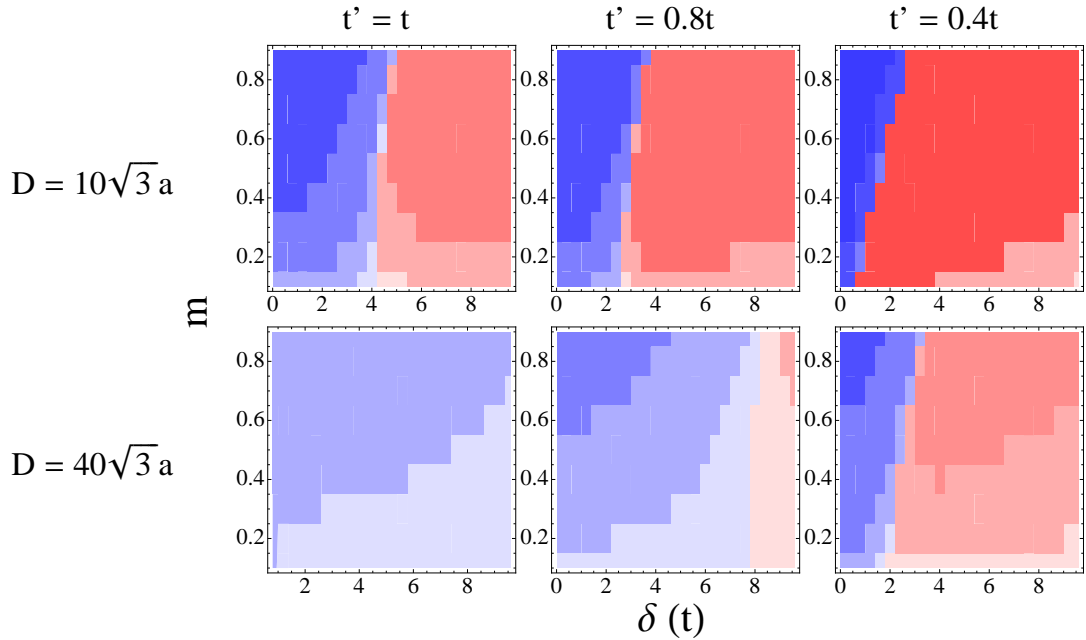}
\caption{$(m, \delta)$ phase-space diagrams for separations of $10 \sqrt{a}$ (top row) and $40 \sqrt{a}$ (bottom row) for three different values of the impurity-carbon hopping parameter $t^\prime$. The sign of the coupling is indicated by the colour (blue for FM and red for AFM) and the strength of the coupling by the degree of shading.}
\label{fig_subsphases} \vspace{-12pt}
\end{figure}

In Figure \ref{fig_subsphases} we present a number of phase diagrams showing the sign and strength of the coupling for different values of these parameters. Each diagram represents an area of $(m, \delta)$ phase space with ferromagnetic (antiferromagnetic) couplings given by a blue (red) shading that is darker for larger magnitude couplings. The diagrams on the top row correspond to a separation of $10 \sqrt{a}$ between the magnetic moments and from left to right show the cases of $t^\prime = 1.0t, 0.8t, 0.4t$. The bottom panels show the same cases for a larger separation of $40 \sqrt{a}$. By examining the border between the blue and red regions in these plots we can infer under what circumstances the sign-change behaviour described above occurs. In all cases the border position varies only weakly with the magnetic moment ($m$) or hence the band-splitting ($V_{ex}$). A stronger dependence is found on the band-centre shift ($\delta$) and we find that, in general, an anti-ferromagnetic alignment is found above
a critical value of $\delta$. The band-centre shift is strongly dependent on the occupation of the magnetic orbital. For a bipartite lattice like graphene a substitutional impurity with a half-filled orbital gives zero band-centre shift when $t^\prime = t$. As we move away from half-filling a larger band-centre shift is required to return the correct band occupation. For smaller values of $t^\prime$ we note that the border between the blue and red regions shifts towards the left, meaning that smaller band-centre shifts will lead to an AFM alignment. Increasing the distance between the impurities reduces the phase-space area corresponding to an AFM alignment by shifting the border to the right and requiring larger band-centre shifts. This finding agrees with the distance-dependent plots in Figure \ref{fig_signchange1} for fixed parameters which show that in the asymptotic limit the coupling changes sign and returns the FM alignment predicted by the RKKY approximation. However, as in the bottom panel of Figure
\ref{fig_signchange1}, the magnitude of the coupling has essentially decayed to zero before the sign change occurs so the only significant coupling between two such impurities is antiferromagnetic. At this point, we note that the theorem proposed by Saremi \cite{saremi:graphenerkky} suggesting that the same (opposite) sublattice coupling is always FM (AFM) only applies for the case of bipartite lattices with half-filling. The configurations discussed here break the half-filling requirement and electron-hole symmetry, at least locally, and this has a dramatic effect on the nature of the interaction. In particular, depending on the moment parameterisation, a strong antiparallel alignment between two substitutional magnetic impurities in graphene may persist to considerable separations. This result is contrary to many predictions made regarding the RKKY interaction in graphene and shows that caution needs to be applied when extrapolating the results of simple calculations to more realistic magnetic impurities.

\vspace {6pt}

\subsection{Multi-Site Impurities}

\begin{figure} [h]
\centering (\textbf{a}) \hspace{0.27\textwidth} (\textbf{b}) \hspace{0.27\textwidth} (\textbf{c})

\hspace{0.5cm}

\includegraphics[width=0.85\textwidth]{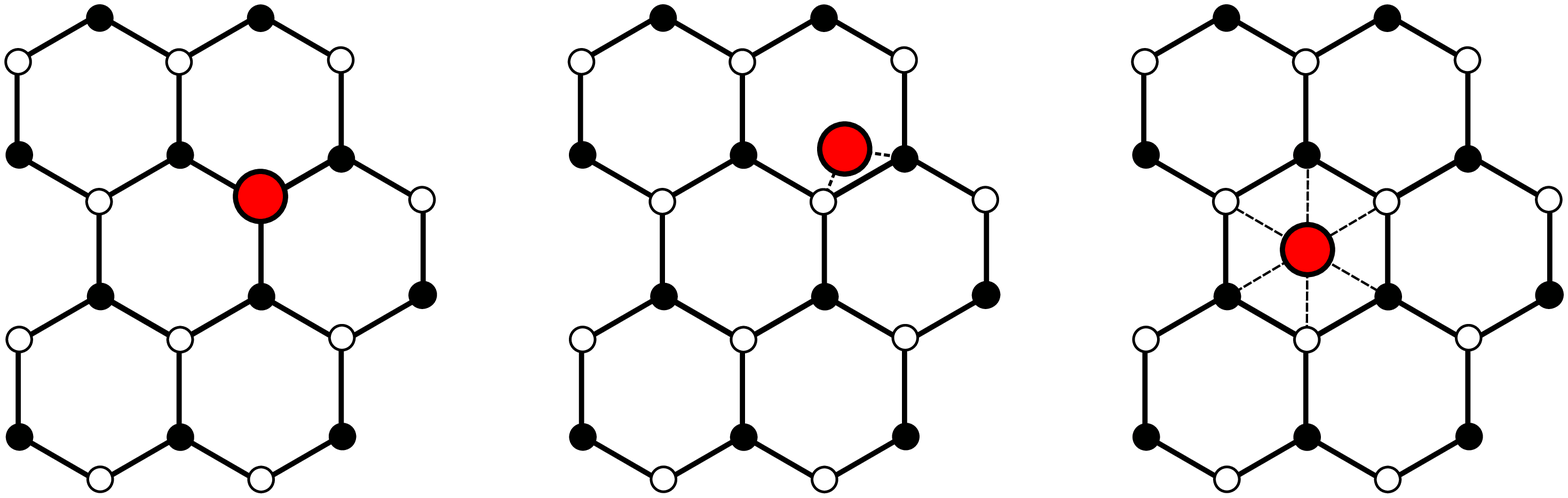}
\caption{The three types of impurity configuration discussed: (\textbf{a}) substitutional atom; (\textbf{b})~bridge-adsorbed atom and (\textbf{c}) plaquette impurity.}
\label{fig_imps}
\end{figure}

The features of the RKKY interaction discussed in Section \ref{Results_sub} contain a strong sublattice dependence. A substitutional or top-adsorbed impurity is associated with a single site on the graphene lattice and therefore also with a single sublattice, as illustrated in Figure \ref{fig_imps}a. We have seen how the sign of the interaction, at least for simple impurity parameterisations, depends on whether the interaction is between sites on the same or on opposite sublattices. However, another type of impurity can be considered which is associated with an equal number of sites on both sublattices. Two examples of this type of impurity are shown in Figure \ref{fig_imps}. These are the \emph{bridge} impurity (b), adsorbed onto the lattice midway between an atom from each sublattice, and the \emph{plaquette}, or centre-adsorbed, impurity at the centre of hexagon (c) and bonding equally to the six surrounding carbon atoms. Plaquette impurities have been investigated more thoroughly than bridge impurities
and a number of interesting predictions have been made. An important difference is noted between the results for incoherent and coherent type moments~ \cite{saremi:graphenerkky}. For incoherent moments, the calculation essentially consists of adding the contributions of the thirty-six single-site interactions between the associated sites of one moment and the other. Coherent moment calculations are more complete and can be performed, for example, using Equation \eqref{staticJ} with the spin-dependent Green functions of a graphene sheet with the two impurities connected to the graphene sheet using the Dyson equation and an appropriate perturbation potential. Saremi noted a decay rate of $D^{-3}$ for incoherent plaquette impurities \cite{saremi:graphenerkky}, a result confirmed by further studies \cite{brey:graphenerkky, black:graphenerkky, sherafati:graphenerkky}. However, Saremi also noted that for the coherent case the $D^{-3}$ term vanishes and that the decay in this instance is faster. This is similar to
the decrease in the interaction range reported for plaquette impurities in metallic carbon nanotubes \cite{David:IEC}, where the decay rate was found to be $D^{-5}$ compared to the $D^{-1}$ rate predicted for substitutional or top-adsorbed impurities. Uchoa \emph{et al.} have since reported a decay rate of $D^{-7}$ for plaquette impurities in graphene \cite{uchoa:rkkygraphene}.

The sign of the interaction between plaquette impurities is also of interest. An always AFM coupling is widely reported in the literature \cite{saremi:graphenerkky, black:graphenerkky, sherafati:graphenerkky}, although Uchoa \emph{et al.} report a FM coupling for smaller values of separation \cite{uchoa:rkkygraphene}. The prevalence of an AFM coupling can be understood from the single-site coupling results. We may initially expect the plaquette impurity to average out the sublattice-driven preference for an always FM or AFM coupling. However, we recall that the magnitude of the AFM coupling in the single site case is three times larger than that of the FM coupling. Thus the AFM contributions from inter-sublattice terms are larger than FM contributions from intra-sublattice terms and an overall AFM coupling is found between plaquette impurities. Another interesting feature for this impurity type is that the period-3 oscillations noted for zigzag separations are no longer present and a smooth monotonic decay
of the coupling is noted for separations in all directions. The jagged features are averaged out by the fact that a plaquette impurity is associated with an equal number of sites from each group defined by the period-3 sequence noted for single-site impurities.

Sherafati and Satpathy \cite{sherafati:graphenerkky} have examined the case of bridge impurities and find a rare example of sign-changing oscillations in the interaction in undoped graphene. They report that every third value of separation has an AFM coupling, and the other two are FM.

\subsection{Spin-Orbit Effects, Electron-Electron Interactions and \emph{ab initio} Calculations}

Having spent some time examining the effects of more detailed impurity parameterisations, it is now worth considering the role a more complicated description of the graphene electronic structure may play. Such a description may include additional terms in a model Hamiltonian calculation or a full \emph{ab~initio} treatment using a Density Functional Theory (DFT) approach. An early study of this type was undertaken by Dugaev \emph{et al.} \cite{dugaev:rkkygraphene} and investigated the interaction in a graphene system where a small gap was introduced by the inclusion of spin-orbit effects. Using an analytic approach based on the Dirac equation, the prediction of an always FM interaction with a long tail exponential decay was made.

Black-Schaffer \cite{black-schaffer_importance_2010} included the effect of electron-electron interactions in a numerical study of the RKKY interaction. These interactions were included using a mean-field solution of a Hubbard Hamiltonian with a finite onsite Coulomb repulsion term $U$ at each site in the graphene lattice. Similar methods have been used previously to investigate intrinsic magnetism in graphene arising from the spin polarisation of localised states near zigzag edges. A number of interesting effects are noticed as the value of $U$ is increased. The oscillatory features in the zigzag direction are removed, as is the factor of three amplitude difference between inter- and intra-sublattice couplings. Furthermore, the decay exponent decreases from $3$ for $\tfrac{U}{t}=0$ to below 2 when $\tfrac{U}{t}>2$. At this value of $U$ the interaction is also isotropic, with no difference between the interactions for armchair and zigzag separations.

A few \emph{ab initio} studies have been performed which address exchange couplings between localised magnetic moments in graphene. These studies are generally difficult to compare with the other results presented here for a number of reasons. Due to a much larger computational cost, \emph{ab initio} calculations are generally only performed for very small values of separation. The majority of these calculations use periodic boundary conditions, which can result in magnetic interactions between impurities in neighbouring unit cells \cite{rapidcomm:emergence, zanolli:vac_defects}. In fact, such interactions can lead to unusual results in \emph{ab initio} studies examining a single magnetic impurity due to the non-oscillatory nature of the RKKY interaction in graphene. Preferential AFM alignment between impurities in neighbouring unit cells can potentially lead to the total suppression of the magnetic moment \cite{rapidcomm:emergence, arkady_embedding_2009}. Periodic unit cell calculations within a \linebreak 
tight-binding scheme can minimise this effect by including a large ``padding'' region around the impurities \cite{black:graphenerkky}, but this is usually not feasible in DFT. Signatures of the RKKY interaction can also be seen in DFT calculations when the effective distance between moments in single impurity concentrations is varied by increasing the size of the unit cell \cite{Lehtinen_diffusionPRL, santos_first-principles_2010}. From a study of pairs of localised moments arising from simple vacancy defects, Pisani \emph{et al.} \cite{pisani_rkkydft} extracted a decay rate of $D^{-1.43}$ for moment separations of up to $25$\AA. This agrees with Black-Schaffer's calculations including electron-electron interactions when the onsite Coulomb repulsion term is $\tfrac{U}{t}=2.1$. A study by Santos \emph{et al.}~\cite{santos_magnetism_2010} however extracts a decay rate of $D^{-2.43}$ for a FM decay between substitutional cobalt impurities on the same sublattice. This result is significantly closer to the $D^{-3}
$
rate predicted without electron- electron interactions. Calculations between Fe atoms meanwhile suggest a preferred AFM alignment between impurities on the same sublattice \cite{rapidcomm:emergence}, suggesting that the nature of the interaction depends strongly on the magnetic species chosen, in agreement with the findings of Section \ref{Results_complicated} of the current work. Moments arising in hydrogenated armchair-edged graphene nanoribbons generally preferred an AFM alignment~\cite{soriano_hGNRs}. Carbon adatoms, adsorbed in the bridge configuration, were found for small separations to have a FM alignment. For larger separations both FM and AFM alignments were found for different values of separation \cite{arkady_rkkydft}.

These findings show that there is still quite a large amount of discrepancy between the interactions predicted using simple RKKY-like models and more complete DFT calculations, which appear very strongly dependent on impurity configuration. Furthermore, direct comparison of results from both methods is often thwarted by the separation constraints imposed by DFT calculations and the possibly oversimplistic treatment of magnetic impurities in model calculations.

\section{Manipulating the Interaction}
\label{Results_manipulation}

In this section we discuss a number of factors which may allow manipulation of magnetic interactions in graphene. A degree of control over the properties of the interaction may be required for spintronic applications, for example to switch on and off spin currents or to dynamically change the magnetic ordering in a system. Another important consideration is that the interactions discussed so far have been very short-ranged, due to a fast decay rate arising from the peculiar electronic dispersion relation in graphene. Methods of augmenting the interaction, or slowing the rate of decay, may thus help overcome some of the difficulties in realising magnetically-doped graphene devices.

\subsection{Geometry Effects}

A number of studies have looked at magnetic interactions in graphene systems with geometries other than the standard two-dimensional sheets considered so far in this work. These include nanotubes~\cite{AntonioDavidIEC, David:IEC, DavidSpinValve, David:PhD, DynamicNJP, bunder:cntrkky}, nanoribbons \cite{bunder:rkkygraphene, black:graphenerkky, black-schaffer_importance_2010, soriano_hGNRs, szalowski2012indirect, sun2012indirect}, nanoflakes \cite{PhysRevB.84.205409} and multilayer graphene systems \cite{hwang:rkkygraphene, kogan:rkkygraphene, killi:rkkybilayer, belonenko:rkkybilayer, jiang:rkkybilayer}. Quasi-one-dimensional graphene systems, such as nanotubes and nanoribbons, are expected to display longer ranged interactions than those present in two-dimensional graphene sheets. Indeed, metallic nanotubes have been shown to display a long ranged $D^{-1}$ interaction when substitutional, top-adsorbed or bridge-adsorbed impurities are considered \cite{AntonioDavidIEC, David:IEC}. However, similar to the case
of graphene, the interaction between plaquette impurities is found to decay at the much faster rate of $D^{-5}$. The sublattice dependent sign rules discussed for graphene also hold for nanotubes within the RKKY approximation, but can be broken when the parameterisation of the impurities is altered \cite{DavidSpinValve}. The eigenstates of carbon nanotubes are subject to periodic boundary conditions in the circumferential direction which ensure the equivalence of all lattice sites. This is not the case in nanoribbon systems where edges are present and each lattice site can be characterised by its distance from the edges. Thus we should expect the interaction to contain a dependence on the location of the impurities across the width of the ribbon. Such a dependence is found in many properties of doped ribbons, including binding energies \cite{me:impseg}, conductance \cite{mucciolo:graphenetransportgaps, Gorjizadeh:defectsribbons} and the magnitude of magnetic moments \cite{me:magprof}. The presence of
localised edge states for zigzag edge geometries \cite{Nakada:1996ribbons, Nakada:1996ribbons2, Son:halfmetallic} should also be expected to introduce features not present in graphene sheets or nanotubes. Black-Schaffer \cite{black:graphenerkky} examined the interaction in nanoribbons, finding that armchair edges did not significantly alter the coupling from the bulk case. However, for localised moments at the edge of zigzag ribbons, an exponentially decaying interaction was reported. The interaction was FM for moments along one edge and AFM for opposite edges. However, the magnitudes of the FM and AFM interactions were equal, unlike the larger AFM magnitude found for the bulk case. Towards the ribbon centre, the results for the bulk were recovered. Including electron-electron interactions in the calculation suggested that the distance dependence vanished for reasonable values of $U$, leading to an extremely long-ranged interaction \cite{black-schaffer_importance_2010}. Sza\l{}owski studied the RKKY
interaction in finite graphene nanoflakes, finding that for certain geometries the addition of a single additional charge carrier could change the sign of the interaction \cite{PhysRevB.84.205409}. A recent work by Klinovaja and Loss extends the study of the interaction in one-dimensional graphenes to include spin-orbit interactions\cite{loss:so_rkky}. The RKKY interaction has been studied in bilayer graphene systems by several authors \cite{hwang:rkkygraphene, kogan:rkkygraphene, killi:rkkybilayer, belonenko:rkkybilayer}. The quadratic nature of the dispersion relation in bilayers means that many of the features present in monolayer graphene are removed. Another interesting feature is that, for A-B stacking, a plaquette impurity in one layer is embedded above a single graphene lattice site on the other layer. Extending the study to more than two layers, Jiang \emph{et al.} \cite{jiang:rkkybilayer} find a number of different decay rate possibilities, depending on the positions of the two impurities and on 
the number of layers.

\subsection{Doping and Disorder}

The key features of the RKKY interaction when the Fermi energy is shifted away from half-filling are clear from the bottom panels of Figure \ref{fig_cases}. These are namely a longer ranged interaction that decays as $D^{-2}$ and sign-changing oscillations, the periods of which depend on the component of the Fermi wavevector in the direction of separation. The appearance of oscillations is due to the breaking of the commensurability effect between the Fermi wavevector and the lattice spacings. Many of the studies of the IEC in graphene establish these key features in doped or gated graphene, but they have been explored in more detail by Sherafati and Satpathy \cite{sherafati:rkkygraphene2}, where the possibility of controlling the sign of the interaction with a gate voltage is raised. The possibility of controlling the interaction using a gate voltage has also been discussed for bilayer graphene \cite{killi:rkkybilayer}.

Lee \emph{et al.} \cite{disorderedRKKY} examine the effects of nonmagnetic disorder on the RKKY interaction in monolayer graphene. The Anderson model was employed to study both onsite and hopping term disorder. Onsite disorder was found to induce sign-changing oscillations in the coupling in certain directions, whereas off-diagonal disorder only effected the amplitude, and not the sign, of the interaction. This suggests that sign-changing behaviour is introduced by the breaking of sublattice symmetry, in a similar manner to that discussed for impurity parameterisation in Section \ref{Results_complicated}. For weak disorder, the interaction retained similar decay behaviour to the clean case. In the strongly disordered, localised regime exponential suppression of the coupling with increasing moment separation was observed. In a recent work, the same authors examine the interplay of disorder and gate voltage \cite{lee2012rkky_preprint}.

\subsection{Strained Graphene Systems}

Strain-engineering of graphene systems has received a lot of attention in recent literature due to the possibility of tuning many of the physical properties of graphene \cite{pereira_tight-binding_2009, Pereira09, Ribeiro09, Pereira10,Guinea:gapsgraphene, Pellegrino11, Klimov12, Mohiuddin09, Levy30072010, Santos12, Peng20123434, me:strainpreprint}. The degree of tuning is enhanced by the different types of strain that can be applied. Apart from simple uniaxial strains \cite{pereira_tight-binding_2009, Ni08}, more exotic features like creases and bubbles can be introduced \cite{Pereira10, Levy30072010, Klimov12, Neekamal12, Xu12, Neekamal12b}. In a recent work, we have explored the possibility of manipulating the IEC in graphene by the application of uniaxial strains parallel or perpendicular to the moment separation direction \cite{me:strainpreprint}. Another recent work also explores this topic \cite{Peng20123434}. We find that both amplification and suppression of the magnetic coupling can be achieved. We
reported a general trend of amplification for strain perpendicular to the moment separation direction, or suppression for strain parallel to this direction. Also noted are oscillations in the amplification as strain is increased for moments separated in non-armchair directions. Such oscillations suggest the intriguing possibility of selectively turning on or off the coupling between moments and also of controlling the inter- and intra-sublattice couplings independently. Since the IEC underpins physical features, including overall moment formation and magnetotransport response, the ability to fine tune the coupling with strain may lead to interesting spintronic applications. Thorough investigation of graphene systems with magnetic impurities and with different types of strain applied may yield a further range of tuneable spintronic properties.

\subsection{Spin Precession and Dynamic RKKY}

One of the major obstacles in the path of exploiting the RKKY interaction between magnetic impurities in graphene for spintronic devices is the very short range of the interaction. It has been suggested that the range of RKKY-like interactions can be augmented if the magnetic moments are set in motion, for example, if they are set to precess by the application of a suitable time-dependent magnetic field \cite{Simanek_gilbert_2003, heinrich_dynamic_2003}. Such an interaction is driven by non-equilibrium spin currents emanating from the precessing moments \cite{filipe:spincurrent}. The magnitude of the interaction can be measured by a quantity called the dynamic spin susceptibility, which describes the response of the magnetism of the system to a dynamic magnetic perturbation. Investigation of this quantity in carbon nanotubes \cite{DynamicNJP} has revealed a decay rate slower than that of the static coupling. Further studies of spin dynamics in graphene systems have suggested the use of these materials as
spin waveguides \cite{filipe:CNTwaveguide}, spin-pumping transistors \cite{filipe:transistor} and spin current lenses \cite{filipe:spinlens}. Recent experimental evidence also suggests possible long-range spin current behaviour in graphene \cite{abanin_giant_2011}. In a recent study \cite{DynamicRKKY}, we examined the spin susceptibility in graphene as a dynamic analogue of the static RKKY coupling, with a particular focus on the separation dependence of the interaction. We find that the decay rate for the dynamic RKKY approaches $D^{-1}$ for both doped and undoped systems, giving a significant augmentation of the interaction range. Furthermore, the dynamic interaction can be related to fluctuations in the lifetimes of magnetic excitations at the impurity sites. Such quantities have been probed in other materials using inelastic scanning tunnelling spectroscopy \cite{heinrich_single-atom_2004, loth_controlling_2010, khajetoorians_detecting_2010, khajetoorians_itinerant_2011}, suggesting that this method may
be used to detect signatures of the dynamic RKKY interaction in graphene.

\section{Conclusion and Experimental Considerations}

In this review we have examined many aspects of indirect exchange interactions in graphene systems. The various approaches to calculating the coupling, ranging from analytical solutions of the RKKY approximation using the Dirac approximation, to numerically expensive \emph{ab initio} total energy calculations, via tight-binding energy difference calculations with varying levels of impurity parameterisation, were discussed. Each of these methodologies has merits and pitfalls, and the contrasting results suggest that there is still some work to be done before useful comparison can be made with experimental results. On the experimental side, it is understandable that with a decay rate as fast as $D^{-3}$ it is difficult to probe the interaction for any reasonable separation. The presence of magnetism in disordered graphene systems may indicate the presence of an exchange coupling between magnetic moments formed around defects. Nuclear magnetic resonance experiments reveal that these defects have indeed magnetic
moments, since they couple to implanted Fe atoms \cite{sielemann_magnetism_2008}. However, whether or not these moments couple with each other, or with the graphene lattice, to form a ferromagnetic state is a controversial subject and many of the results in this area have proved difficult to reproduce~\cite{yazyev:review}. Progress in spin-dependent scanning tunnelling spectroscopy has enabled the RKKY interaction to be measured directly from the magnetisation curves of individual magnetic adatoms \cite{meier2008revealing, zhou2010strength}. As discussed above, signatures of the dynamic analogue of the RKKY interaction may be detectable from the magnetic excitation lifetimes of individual moments measured using inelastic scanning tunnelling spectroscopy \cite{heinrich_single-atom_2004, loth_controlling_2010, khajetoorians_detecting_2010, khajetoorians_itinerant_2011}. In this case, the longer range of the interaction may make experimental detection more feasible \cite{DynamicRKKY}. Using a gate voltage to
increase the interaction range may also be necessary to probe the interaction in the static case. To our knowledge these types of experiment have not yet been performed with a graphene host medium. Signatures of indirect exchange interactions may also be present in transport measurements taken in magnetically-doped graphene systems \cite{vojislav:acs, DavidSpinValve}, since different orientations of the magnetic moments lead to different scattering regimes for up- and \linebreak down-spin electrons.

Future work on the topic of indirect exchange interactions in magnetically doped graphene should focus on the gaps and discrepancies between the various theoretical models, and between these models and possible experimental configurations. In particular, the strength of the interaction for specific magnetic impurities and whether it survives from the ideal model to experimentally realisable conditions of finite temperature and lattice disorder are important considerations. Another area of exploration is the interplay between magnetic impurities and other features that can be introduced into the graphene lattice. The topics of different geometries, in the form of nanoribbons and nanotubes, and of geometrical distortions, in the form of strain, were briefly discussed in this review and results were mentioned for simple cases. More complicated geometries or distortions may introduce new features and more possibilities to manipulate the interaction. In moving towards technological application of RKKY-like
interactions, the study of larger ensembles of magnetic impurities will be necessary. The magnetic and electronic properties that emerge from such ensembles should be dictated by the pairwise indirect exchange interactions discussed in this review, and may lead to a range of interesting features and~applications.

\section*{Acknowledgements}\\

The authors acknowledge financial support received from the Irish Research Council for Science, Engineering and Technology under the EMBARK initiative and from Science Foundation Ireland under Grant No. SFI 11/RFP.1/MTR/3083. The Center for Nanostructured Graphene CNG is sponsored by the Danish National Research Foundation.

\bibliographystyle{mdpi}
\makeatletter
\renewcommand\@biblabel[1]{#1. }
\makeatother

\begin{thebibliography}{------}
\bibitem[qua()]{quantisation_footnote}
The allowed electron energies are quantised in one-dimensional or multilayer systems. In higher dimensions the allowed energies are modified by the wells, but are not strictly quantised.
\end{thebibliography}

%
%
%

\end{document}